**Article Title**

ToF-SIMS data analysis of *Shewanella oneidensis* MR-1 biofilms

**Authors**

Gabriel D. Parker[1], Andrew Plymale[2], Luke Hanley[1], and Xiao-Ying Yu[3*]
**Affiliations**
1. Department of Chemistry, University of Illinois Chicago, Chicago, IL 60607.
2. Energy and Environment Directorate, Pacific Northwest National Laboratory, Richland, WA 99352.
3. Materials Science and Technology Division, Oak Ridge National Laboratory, Oak Ridge, TN 37380.
**Corresponding author(s)**
Xiao-Ying Yu (yuxiaoying@ornl.gov)

**Abstract**

Analysis of bacterial biofilms is particularly challenging and important with diverse applications from systems biology to biotechnology. Among the variety of techniques that have been applied, time-of-flight secondary ion mass spectrometry (ToF-SIMS) has many promising features in studying the surface characteristics of biofilms. ToF-SIMS offers high spatial resolution and high mass accuracy, which permit surface sensitive analysis of biofilm components. Thus, ToF-SIMS provides a powerful solution to addressing the challenge of bacterial biofilm analysis. This dataset covers ToF-SIMS analysis of *Shewanella oneidensis* MR-1 isolated from freshwater lake sediment in New York state. The MR-1 strain is known to have metal and sulfur reducing properties and it can be used for bioremediation and wastewater treatment. There is a current need to identify small molecules and fragments produced from bacterial biofilms. Static ToF-SIMS spectra of MR-1 were obtained using an IONTOF TOF.SIMS V instrument equipped with a 25 keV $Bi_3^+$ metal ion gun. Identified molecules and molecular fragments are compared against known biological databases and the reported peaks have at least 65 ppm mass accuracy. These molecules range from lipids and fatty acids to flavonoids, quinolones, and other naturally occurring organic compounds. It is anticipated that the spectral identification of key peaks will assist detection of metabolites, extracellular polymeric substance molecules like polysaccharides, and biologically relevant small molecules using ToF-SIMS in future surface and interface research.

**Keywords**

**Specifications Table**

| Subject | Organic surfaces and interfaces |
|---|---|
| Specific subject area | *Shewanella oneidensis MR-1* biofilm surface analysis |
| Type of data | Table<br>Figure |
| How data were acquired | Mass Spectrometry<br>Instruments: Time-of-Flight Secondary Ion Mass Spectrometry<br>Make and model and of the instruments used: IONTOF TOF.SIMS V |
| Data format | Raw<br>Analyzed |
| Parameters for data collection | 60 scans with 25 keV $Bi_3^+$ metal ion gun rastering over a 500 × 500 μm$^2$ area used with oxygen flood gun for charge compensation. |
| Description of data collection | Static ToF-SIMS spectra were obtained using an IONTOF TOF.SIMS V equipped with a 25 keV $Bi_3^+$ metal ion gun. The raster size for each region of interest was 500 × 500 μm$^2$. The number of scans per spectrum is 60. Spectra requiring use of the flood gun were specified for each sample. Peak identifications are based on mass formula with a mass deviation of less than 65 ppm. |
| Data source location | Institution: Oak Ridge National Laboratory<br>City/Town/Region: Oak Ridge, TN<br>Country: United States of America |
| Data accessibility | Available via Dr. Xiao-Ying Yu's GitHub site, git@code.ornl.gov:nsrd_au/data_pub_sharing.git |

**Value of the Data**

- *Shewanella* MR-1 strains, have been shown to reduce metals such as iron, uranium, and other metals by using oxygen, metal oxides, nitrates, and sulfates as electron acceptors [1]. These properties can affect secretions and synthesis of metabolites and other chemicals that are building blocks of biofilms. Understanding and identifying small molecule production that a bacterial strain possesses is crucial since there is no existing database for biofilms and components analysed by ToF-SIMS.

- Using specific bacteria strains to enhance bioremediation is desirable for industries, like agriculture, nuclear waste management, and health care. Identifying compounds bacterial strains secrete can lead to advancement in diverse applications and support efforts towards building a community database for identification of small molecules inherent of extracellular polymeric substance (EPS) via ToF-SIMS.

- This dataset provides identification of small organic molecules, amino acids, fatty acids, and lipids among other compounds for *Shewanella oneidensis*. This data can be used as control datasets for the specified organism as a basis for identification of molecules in both positive and negative modes.

**Data Description**

We focus on reporting the molecules with relevance to EPS [2, 3]. Collected ToF-SIMS spectra show the mass range of 0-800 Da while highlighting the mass range 150-500 Da, as this region is of interest for metabolite, fatty acid, and lipid production among other molecules. **Figure 1** shows the ToF-SIMS spectra of *Shewanella oneidensis* MR-1 (MR-1) captured in the negative mode. **Table 1** shows possible peak identifications for the MR-1 biofilm in the negative mode. **Figure 2** shows the spectra of *Shewanella oneidensis* MR-1 acquired in the positive mode. **Table 2** shows possible peak identifications in the positive mode. Each mode provides complementary information of the bacteria. We highlight molecules, such as amino acids, phosphates, sulfates, fatty acids, and lipids, that are important EPS components of the MR-1 bacterial system. Supplementary **Figures S1** and **S2** depict ToF-SIMS spectral plots of MR-1 planktonic cells. Similarly, Supplementary **Tables S1 – S6** show ToF-SIMS acquisition conditions used for biofilm and planktonic samples and key peak identification of MR-1 planktonic cells, respectively.

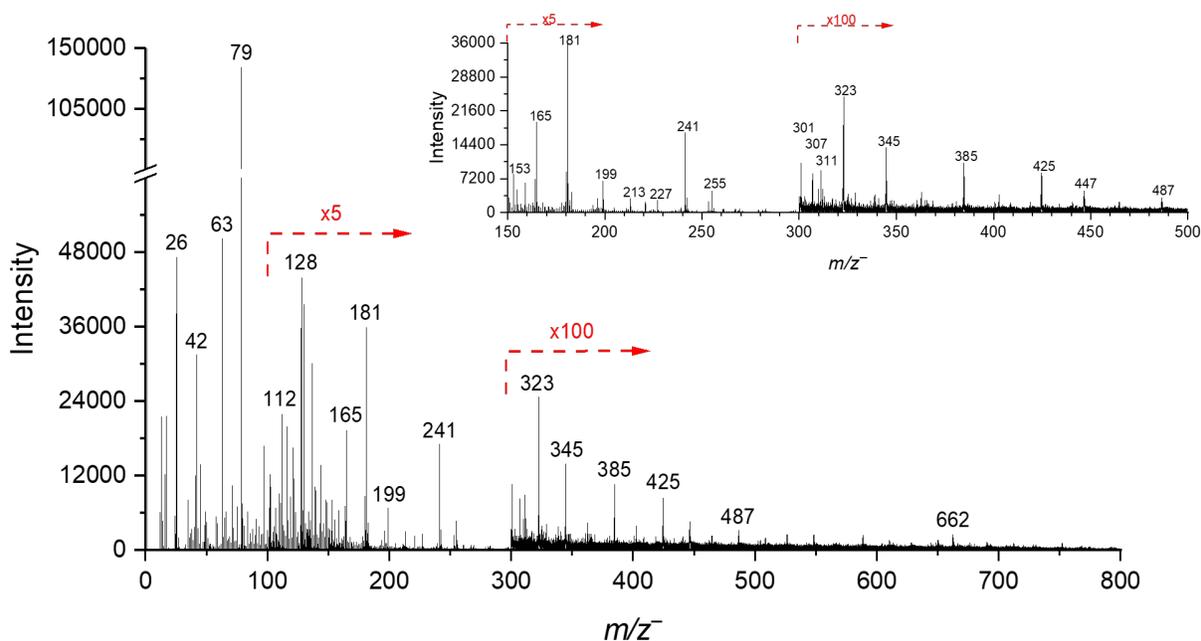

**Figure 1:** Static ToF-SIMS spectra of *Shewanella oneidensis* biofilms in the negative mode showing characteristic fatty acids, lipids, and amino acids.

**Table 1:** Possible peak identifications of *Shewanella oneidensis* biofilms in the negative ion mode.

| $m/z^-_{theo.}$ | $m/z^-_{obs.}$ | ΔM | Species | Assignment | References |
|---|---|---|---|---|---|
| 62.96 | 62.97 | 59.82 | $PO_2^-$ | Hypophosphite | This work |
| 96.97 | 96.97 | 43.92 | $PO_4H_2^-$ | Phosphoric acid | This work |
| 109.98 | 109.97 | 36.89 | $CH_3PO_4^-$ | Formylphosphonate | This work |
| 111.95 | 111.95 | 6.44 | $SO_5^-$ | Peroxomonosulfate | This work |
| 122.00 | 122.00 | 5.53 | $C_2H_5PNO_3^-$ | Amidophosphate | This work |
| 129.08 | 129.09 | 43.97 | $C_6H_{11}NO_2^-$ | Amino acid | [4] |
| 153.01 | 153.01 | 5.48 | $C_3H_7NO_4S^-$ | Amino acid | This work |
| 158.93 | 158.93 | 38.11 | $P_2O_6H^-$ | Hypophosphate | This work |
| 180.04 | 180.05 | 29.40 | $C_9H_8O_4^-$ | Cinnamic acid | This work |
| 199.17 | 199.17 | 17.02 | $C_{12}H_{23}O_2^-$ | Fatty acid | [5] |
| 213.19 | 213.18 | 12.70 | $C_{13}H_{25}O_2^-$ | Fatty acid | This work |
| 227.20 | 227.20 | 1.52 | $C_{14}H_{27}O_2^-$ | Fatty acid | [5] |
| 241.22 | 241.22 | 10.91 | $C_{15}H_{29}O_2^-$ | Fatty acid | [5] |
| 255.23 | 255.23 | 9.69 | $C_{16}H_{31}O_2^-$ | Fatty acid | [5] |
| 260.87 | 260.87 | 3.16 | $Na_3S_2O_8^-$ | Media | This work |
| 276.85 | 276.86 | 34.67 | $Na_3S_3O_7^-$ | Media | This work |
| 311.17 | 311.16 | 15.25 | $C_{17}H_{27}SO_3^-$ | Benzenesulfonic acids | This work |
| 325.18 | 325.18 | 3.38 | $C_{18}H_{29}SO_3^-$ | Benzenesulfonic acids | This work |
| 548.41 | 548.43 | 37.17 | $C_{33}H_{56}O_6^-$ | Sterol lipid | This work |
| 662.45 | 588.52 | 27.80 | $C_{38}H_{68}O_4^-$ | Glycerolipid | This work |
| 62.96 | 662.47 | 24.20 | $C_{36}H_{71}O_8P^-$ | Glycerophosphates | This work |

Footnotes:

$m/z^-_{theo.}$: theoretical mass to charge ratio in the negative ion mode.

$m/z^-_{obs.}$: observed mass to charge ratio in the negative ion mode.

$\Delta M = Abs (10^6 \times (m/z^-_{obs.} - m/z^-_{theo.})/ m/z^-_{theo.})$ (expressed in ppm) [6].

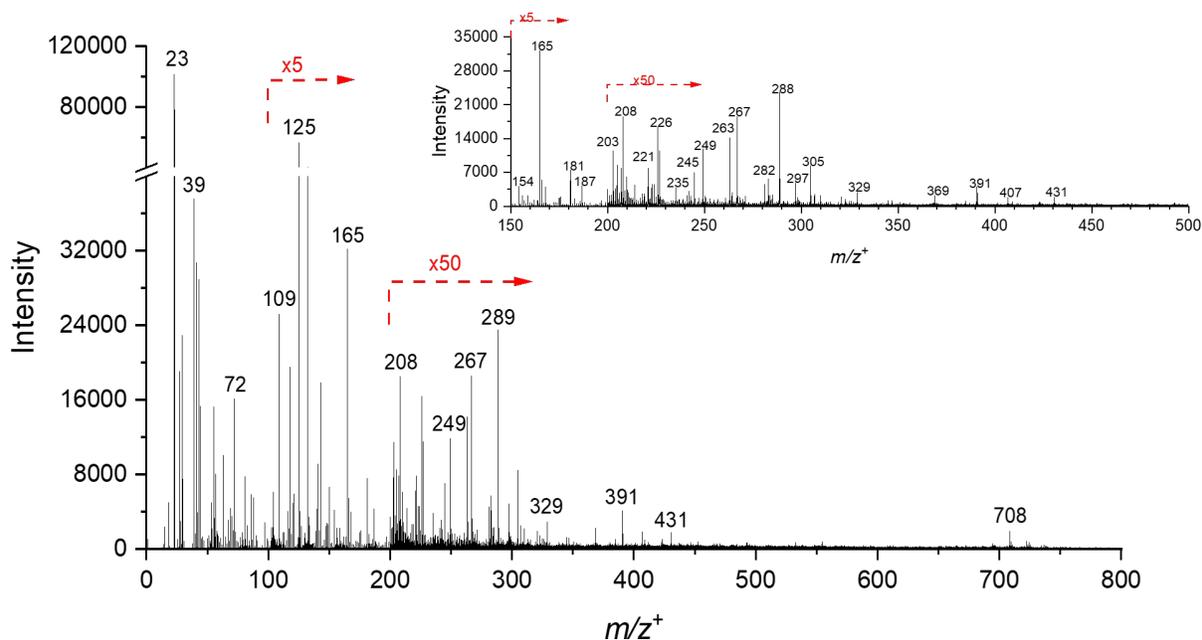

**Figure 2:** Static ToF-SIMS spectra of *Shewanella oneidensis* biofilms in the positive mode, showing characteristic fatty acids, lipids, and amino acids.

**Table 2:** Possible peak identifications of *Shewanella oneidensis* biofilm s in the positive ion mode.

| $m/z^-_{theo.}$ | $m/z^-_{obs.}$ | ΔM | Species | Assignment | References |
|---|---|---|---|---|---|
| 108.94 | 108.95 | 58.60 | $Na_2PO_2^+$ | Media | This work |
| 118.10 | 118.10 | 1.00 | $C_6H_{14}O_2^+$ | Ethers | This work |
| 132.11 | 132.11 | 26.50 | $C_5H_{14}N_3O^+$ | Alkaloids | This work |
| 154.09 | 154.09 | 18.81 | $C_8H_{12}NO_2^+$ | Phenylethylamines | This work |
| 164.93 | 164.94 | 49.85 | $Na_3PO_4H^+$ | Media | This work |
| 176.07 | 176.07 | 0.57 | $C_{10}H_{10}NO_2^+$ | Coumarins | This work |
| 180.92 | 180.92 | 0.50 | $Na_3SO_5^+$ | Media | This work |
| 186.91 | 186.92 | 29.89 | $Na_4PO_4^+$ | Media | This work |
| 196.90 | 196.89 | 61.56 | $NaP_2O_7^+$ | Media | This work |
| 226.05 | 226.04 | 48.80 | $C_{10}H_{10}O_6^+$ | Dicarboxylic acids | This work |
| 235.17 | 235.17 | 10.14 | $C_{15}H_{23}O_2^+$ | Isoprenoids | This work |
| 249.18 | 249.19 | 27.61 | $C_{11}H_{25}N_2O_4^+$ | Dipeptide | This work |
| 263.20 | 263.20 | 12.82 | $C_{17}H_{27}O_2^+$ | Fatty acid | This work |
| 281.15 | 281.16 | 26.26 | $C_{19}H_{21}O_2^+$ | Lignans | This work |
| 282.85 | 282.86 | 19.51 | $Na_2P_3O_9^+$ | Media | This work |
| 288.86 | 288.87 | 33.28 | $Na_5P_2O_7^+$ | Media | This work |
| 297.21 | 297.19 | 51.51 | $C_{17}H_{29}O_4^+$ | Fatty acid | This work |
| 368.83 | 368.84 | 41.77 | $Na_5P_3O_{10}H^-$ | Media | This work |
| 532.52 | 532.54 | 32.36 | $C_{36}H_{68}O_2^+$ | Wax monoesters | This work |

Footnotes:

$m/z^-_{theo.}$: theoretical mass to charge ratio in the negative ion mode.
$m/z^-_{obs.}$: observed mass to charge ratio in the negative ion mode.
ΔM = Abs ($10^6$ × ($m/z^-_{obs.}$ − $m/z^-_{theo.}$)/ $m/z^-_{theo.}$) (expressed in ppm) [6].

**Experimental Design, Materials, and Methods**

The *Shewanella* MR-1 bacteria strain was cultured on Tryptic Soy Broth (TSB) agar plates [7]. All agar plates were incubated at 30 °C for 24 hrs. After assessing cultures for contamination, one to two pure colonies were inoculated in 10 mL of TSB medium. Planktonic cells were grown to ~0.6 optical density ($OD_{600}$) [5, 7]. Planktonic cells were then harvested by centrifugation for 5 min. at 5000 rpm. After centrifugation, the supernatant was discarded and replaced with 1 mL sterile deionized (DI) water to resuspend. This process was repeated three times, then 200 μL DI water was added for a final resuspension. The planktonic cells were then plated onto sterilized silicon (Si) wafers and air dried under laminar flow within a biosafety cabinet (BSC) [8]. Biofilms were cultured using static cells described previously [4]. Biofilms were grown for 5-6 days, and maturation was observed via optical microscope. Biofilms were then harvested onto sterilized Si wafers and dried under laminar flow of nitrogen within a BSC. Si wafer controls were prepared by sonication of wafers in 30 mL ethanol, isopropanol, and acetone, respectively for 5 min. The wafers were then blown dry with nitrogen gas after sonication. The cleaned Si wafers were then treated with UV-ozone (model No. 342, Jetlight Company Inc.) for 1 min to render the surface hydrophilic [4, 9].

Static ToF-SIMS spectra were obtained using an IONTOF TOF.SIMS V equipped with a 25keV $Bi_3^+$ metal ion gun. The raster size for each region of interest was 500 × 500 μm$^2$. The number of scans per spectra is 60. **Tables 1** and **2** showed peak identification with suggested mass formula, and each peak has a mass deviation of less than 65 ppm. The SIMS mass accuracy is defined as $\Delta M := \mathrm{Abs}\ (10^6 \times (m/z_{obs} - m/z_{theo})/\ m/z_{theo})$ (expressed in ppm), where $m/z_{obs}$ and $m/z_{theo}$ refer to the observed and theoretical mass to charge ratio of a specific peak in the negative or positive ion mode [6, 10].

The identification procedure followed a multi-step process. The IonTOF database embedded in the SurfaceLab software (version 7.2) is not used, while the IonTOF peak searching and mass matching functions are used [11]. The mass matching function calculates different combinations of periodic table elements to mass match a selected peak with a mass deviation and match score. These combinations range from organics to inorganics. Using the "peak search" function within *SurfaceSpectra* functions, with parameters of SNR 3.0, max background 0.8, and minimum counts 25, a peak list was generated giving the best mass matching formula for the given *m/z* value under each specific spectrum.

Peak assignment was further verified by literature search pertaining to these molecules. Many databases were surveyed for molecular assessment such as PubMed, ChEBI, LipidMaps, MetabolomicsWorkbench, and KEGG. Assignment was based on the mass accuracy from the associated peak value to the corresponding value within the aforementioned databases. All identifications are molecular classes only due to the large number of possibilities each *m/z* value presents, with the best match based on relative mass accuracy. Major species identified here are amino acids, fatty acids, lipids, and other relevant molecules, this finding is in agreement with previous results [12, 13]. More specific identification of molecules would need to be corroborated with tandem MS and/or other complimentary techniques [14, 15]. This approach of peak identification also applies to *m/z* values below 150. ToF-SIMS spectral results presented here are useful in future biotechnology and fundamental research of biofilms.

**Acknowledgments**
Gabriel Parker was funded by the Department of Energy SCGSR fellowship program and the Department of Energy GRO internship program at Oak Ridge National Laboratory. The authors would like to thank Dr. Yuchen Zhang Dr. Jiyoung Son, and Dr. Rachel Komorek for their contribution for the sample analysis and data collection. Dr. Xiao-Ying Yu is indebted to the support of the strategic Laboratory Directed Research and Development of the Physical Sciences Directorate of the Oak Ridge National Laboratory (ORNL).

This manuscript has been authored by UT-Battelle, LLC under Contract No. DE-AC05-00OR22725 with the U.S. DOE. The United States Government retains and the publisher, by accepting the article for publication, acknowledges that the United States Government retains a non-exclusive, paid-up, irrevocable, worldwide license to publish or reproduce the published form of this manuscript, or allow others to do so, for United States Government purposes. The DOE will provide public access to these results of federally sponsored research in accordance with the DOE Public Access Plan (http://energy.gov/downloads/doe-public-access-plan).

**Declaration of Competing Interest**

The authors declare that they have no known competing financial interests or personal relationships which have, or could be perceived to have, influenced the work reported in this article.

**Article Title**

ToF-SIMS data analysis of *Shewanella oneidensis* MR-1 biofilms

**Authors**

Gabriel D. Parker[1], Andrew Plymale[2], Luke Hanley[1], and Xiao-Ying Yu[3*]

**Affiliations**

1. Department of Chemistry, University of Illinois Chicago, Chicago, IL 60607.
2. Energy and Environment Directorate, Pacific Northwest National Laboratory, Richland, WA 99352.
3. Materials Science and Technology Division, Oak Ridge National Laboratory, Oak Ridge, TN 37380.

**Corresponding author(s)**

Xiao-Ying Yu (yuxiaoying@ornl.gov)

# Table of Contents

**Additional Figures**

The supplementary figures provided here are additonal to the *Shewanella* MR-1 biofilm as described in the main text. **Figure S1** and **Figure S2** provide insight for the planktonic cell bacteria composition. Differences among the planktonic cells and biofilms are related to the extracellular polymeric substance (EPS) sectretion, which the biofilm employs to congregate and pack planktonic cells together ultimately forming the colony. It is known that metabolites and other compounds secretions differ in the aerobic versus the anaerobic sides of the biofilms.

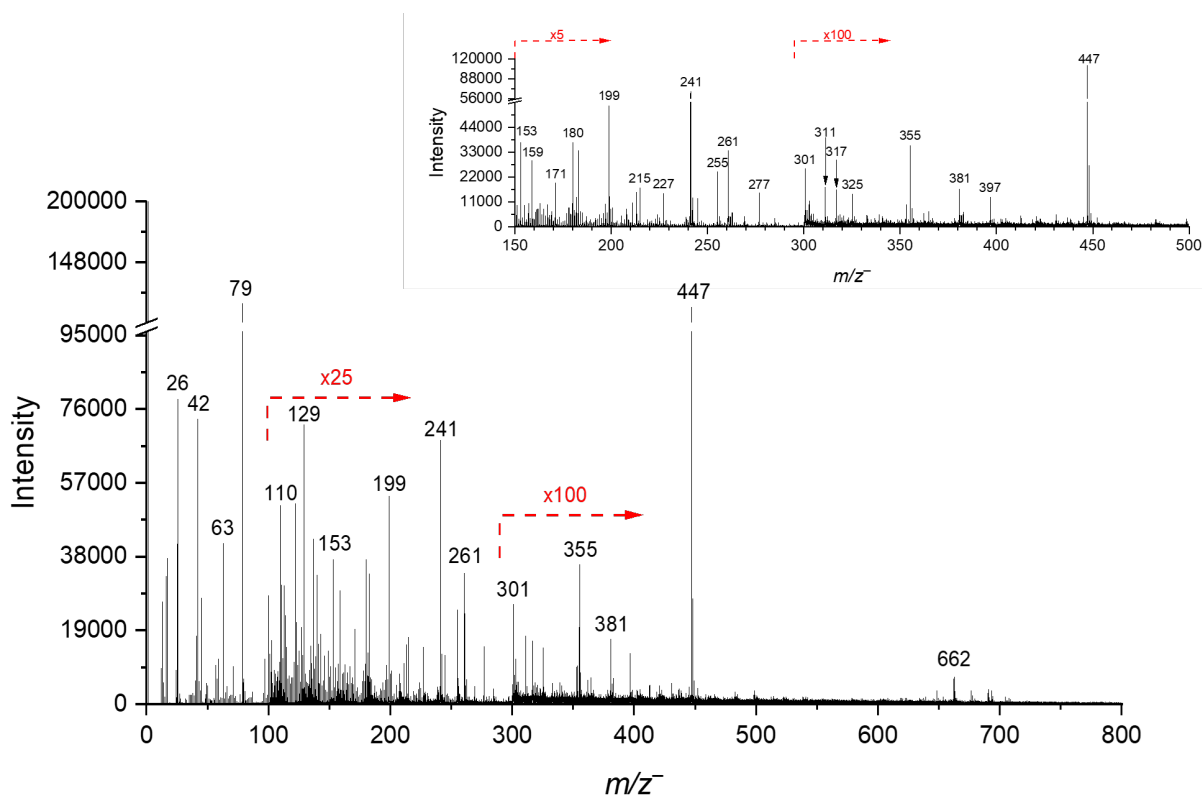

**Figure S1:** Static ToF-SIMS spectra of *Shewanella oneidensis* planktonic cells in the negative ion mode.

**FigureS1** provides more information related to the planktonic cells secretions in the negative mode, identifying characteristic fatty acids, lipids, and amino acids specific to this bacterial species.

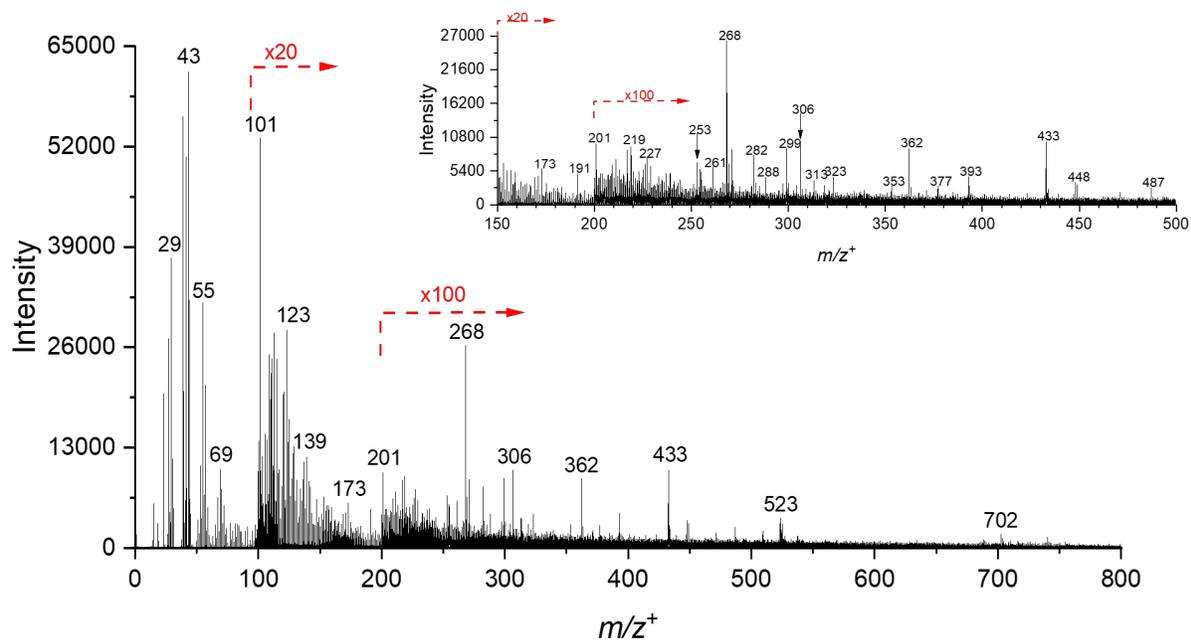

**Figure S2:** Static ToF-SIMS spectra of *Shewanella oneidensis* planktonic cells in the positive mode.

**FigureS2** provides more information related to the planktonic cells secretions in the positive mode, identifying characteristic fatty acids, lipids, and amino acids specific to this bacterial species.

**Additional Tables**

The supplementary tables provided here cover the instrument parameters used to collect the spectra observed and other peak identification tables for the planktonic cells **Figures S1** and **S2**.

**Table S1:** ToF-SIMS spectral acquisition parameters used for *Shewanella oneidensis biofilm* in negative ion mode.

| *Sample* | ***Shewanella oneidensis* biofilm** |
|---|---|
| *Instrument* | IONTOF ToF-SIMS V |
| *Mass Analyzer* | Time-of-Flight |
| *Analyzer Mass Resolution* | 3193 m/Δm |
| *Mass Resolution m/z value* | 59.0173; $C_2H_3O_2^-$ |
| *Calibration Species* | $CH_2^-$, $CHO_2^-$, $C_2H_2NO_2^-$, $C_4H_6NO_2^-$, $C_{15}H_{29}O_2^-$, $C_{16}H_{31}O_2^-$ |
| *Primary Ion Gun* | Liquid Metal Ion Gun |
| *Primary Ion Source* | $Bi_3^+$ |
| *Primary Ion Dose* | 2.45 x $10^{11}$ ion/cm$^2$ |
| *Pulse Width* | 25.0 ns |
| *Pulse Rate* | 10.0 kHz |
| *Beam Diameter* | 5 μm |
| *Analysis Area* | 500 μm × 500 μm |
| *Polarity* | Negative |
| *Mass Range* | 0-800 |
| *Spectra Acquisition Time* | 98.30 s |

**Table S2:** ToF-SIMS spectral acquisition parameters used for *Shewanella oneidensis biofilm* in positive ion mode.

| Sample | *Shewanella oneidensis* biofilm |
|---|---|
| *Instrument* | IONTOF ToF-SIMS V |
| *Mass Analyzer* | Time-of-Flight |
| *Analyzer Mass Resolution* | 4340 m/Δm |
| *Mass Resolution m/z value* | 27.0231; $C_2H_3^+$ |
| *Calibration Species* | $CH_3^+$, $C_4H_5^+$, $C_8H_{10}NO^+$, $C_{19}H_{21}O_2^+$ |
| *Primary Ion Gun* | Liquid Metal Ion Gun |
| *Primary Ion Source* | $Bi_3^+$ |
| *Primary Ion Dose* | $2.45 \times 10^{11}$ ion/cm$^2$ |
| *Pulse Width* | 25.0 ns |
| *Pulse Rate* | 10.0 kHz |
| *Beam Diameter* | 5 μm |
| *Analysis Area* | 500 μm × 500 μm |
| *Polarity* | Positive |
| *Mass Range* | 0-800 |
| *Spectra Acquisition Time* | 98.30 s |

**Table S3:** Possible peak identifications of *Shewanella oneidensis* planktonic cells in the negative ion mode.

| $m/z^-_{theo.}$ | $m/z^-_{obs.}$ | ΔM | Species | Assignment |
|---|---|---|---|---|
| 62.96 | 62.97 | 48.93 | $PO_2^-$ | Hypophosphite |
| 96.97 | 96.97 | 31.17 | $PO_4H_2^-$ | Phosphoric acid |
| 109.98 | 109.97 | 34.17 | $CH_3PO_4^-$ | Formylphosphonate |
| 122.00 | 122.00 | 5.67 | $C_2H_5PNO_3^-$ | Amidophosphate |
| 129.08 | 129.09 | 48.61 | $C_6H_{11}NO_2^-$ | Amino acid |
| 153.01 | 153.01 | 1.50 | $C_3H_7NO_4S^-$ | Amino acid |
| 158.93 | 158.93 | 26.56 | $P_2O_6H^-$ | Hypophosphate |
| 171.10 | 171.11 | 64.44 | $C_9H_{15}O_3^-$ | Fatty acid |
| 180.04 | 180.05 | 37.89 | $C_9H_8O_4^-$ | Cinnamic acid |
| 198.90 | 198.90 | 16.86 | $NaS_2O_7^-$ | Media |
| 199.17 | 199.17 | 24.45 | $C_{12}H_{23}O_2^-$ | Fatty acid |
| 213.19 | 213.18 | 6.72 | $C_{13}H_{25}O_2^-$ | Fatty acid |
| 227.20 | 227.20 | 5.16 | $C_{14}H_{27}O_2^-$ | Fatty acid |
| 241.22 | 241.22 | 11.72 | $C_{15}H_{29}O_2^-$ | Fatty acid |
| 255.23 | 255.23 | 1.81 | $C_{16}H_{31}O_2^-$ | Fatty acid |
| 276.85 | 276.83 | 64.80 | $Na_3S_3O_7^-$ | Media |
| 311.17 | 311.16 | 18.94 | $C_{17}H_{27}SO_3^-$ | Benzenesulfonic acids |
| 325.18 | 325.18 | 8.62 | $C_{18}H_{29}SO_3^-$ | Benzenesulfonic acids |
| 355.16 | 355.16 | 8.04 | $C_{21}H_{23}O_5^-$ | Flavonoid |
| 447.13 | 447.12 | 21.97 | $C_{22}H_{23}O_{10}^-$ | Flavonoid |

Footnotes:
$m/z^-_{theo.}$: theoretical mass to charge ratio in the negative ion mode.
$m/z^-_{obs.}$: observed mass to charge ratio in the negative ion mode.
ΔM:= Abs ($10^6$ × ($m/z^-_{obs.}$ − $m/z^-_{theo.}$)/ $m/z^-_{theo.}$) (expressed in ppm) [1].

**Table S4:** ToF-SIMS spectral acquisition parameters used for *Shewanella oneidensis* planktonic cells in the negative ion mode.

| Sample | *Shewanella oneidensis* **planktonic cells** |
|---|---|
| *Instrument* | IONTOF ToF-SIMS V |
| *Mass Analyzer* | Time-of-Flight |
| *Analyzer Mass Resolution* | 3852 m/Δm |
| *Mass Resolution m/z value* | 59.0148; $C_2H_3O_2^-$ |
| *Calibration Species* | $CH_2^-$, $CHO_2^-$, $C_4H_6NO_2^-$, $C_{15}H_{29}O_2^-$, $C_{16}H_{31}O_2^-$ |
| *Primary Ion Gun* | Liquid Metal Ion Gun |
| *Primary Ion Source* | $Bi_3^+$ |
| *Primary Ion Dose* | 2.45 x $10^{11}$ ion/cm$^2$ |
| *Pulse Width* | 25.0 ns |
| *Pulse Rate* | 10.0 kHz |
| *Beam Diameter* | 5 μm |
| *Analysis Area* | 500 μm × 500 μm |
| *Polarity* | Negative |
| *Mass Range* | 0-800 |
| *Spectra Acquisition Time* | 98.30 s |

**Table S5:** Possible peak identifications of *Shewanella oneidensis* planktonic cells in the positive ion mode.

| $m/z^-_{theo.}$ | $m/z^-_{obs.}$ | ΔM | Species | Assignment |
|---|---|---|---|---|
| 101.06 | 101.06 | 33.09 | $C_5H_9O_2^+$ | Fatty acids |
| 113.07 | 113.07 | 0.67 | $C_5H_9N_2O^+$ | Deoxythymine |
| 123.08 | 123.08 | 34.09 | $C_8H_{11}O^+$ | Phenols |
| 139.08 | 139.07 | 39.49 | $C_8H_{11}O_2^+$ | Phenols |
| 153.09 | 153.09 | 17.85 | $C_9H_{13}O_2^+$ | Fatty lactones |
| 173.02 | 173.02 | 10.13 | $C_3H_{10}O_6P^+$ | Phosphate esters |
| 191.12 | 191.13 | 52.08 | $C_{11}H_{15}N_2O^+$ | Tryptamines |
| 227.13 | 227.13 | 0.85 | $C_{12}H_{19}O_4^+$ | Jasmonic acids |
| 253.09 | 253.10 | 52.95 | $C_{16}H_{13}O_3^+$ | Flavonoids |
| 268.24 | 268.25 | 40.72 | $C_{17}H_{32}O_2^+$ | Fatty acids |
| 282.26 | 282.26 | 5.86 | $C_{18}H_{34}O_2^+$ | Fatty acids |
| 299.26 | 299.25 | 36.75 | $C_{18}H_{35}O_3^+$ | Fatty acids |
| 306.26 | 306.25 | 22.42 | $C_{20}H_{34}O_2^+$ | Fatty acids |
| 313.27 | 313.26 | 35.84 | $C_{19}H_{37}O_3^+$ | Fatty acids |
| 323.10 | 323.12 | 44.57 | $C_{10}H_{19}N_4O_6S^+$ | Tripeptides |
| 362.28 | 362.27 | 31.71 | $C_{23}H_{38}O_3^+$ | Fatty acid |
| 393.04 | 393.06 | 53.74 | $C_{13}H_{17}N_2O_8S_2^+$ | Beta lactams |
| 433.09 | 433.07 | 45.74 | $C_{24}H_{17}O_8^+$ | Flavonoids |
| 448.10 | 448.11 | 15.77 | $C_{21}H_{20}O_{11}^+$ | Flavonoids |
| 523.47 | 523.45 | 47.50 | $C_{33}H_{63}O_4^+$ | Glycerolipids |
| 702.41 | 702.42 | 9.25 | $C_{36}H_{63}O_{11}P^+$ | Glycerophospholipids |

Footnotes:

$m/z^-_{theo.}$: theoretical mass to charge ratio in the negative ion mode.

$m/z^-_{obs.}$: observed mass to charge ratio in the negative ion mode.

ΔM:= Abs ($10^6$ × ($m/z^-_{obs.}$ − $m/z^-_{theo.}$)/ $m/z^-_{theo.}$) (expressed in ppm) [1].

**Table S6:** ToF-SIMS spectral acquisition parameters used for *Shewanella oneidensis* planktonic cells in the positive ion mode.

| *Sample* | |
|---|---|
| *Instrument* | IONTOF ToF-SIMS V |
| *Mass Analyzer* | Time-of-Flight |
| *Analyzer Mass Resolution* | 3877 m/Δm |
| *Mass Resolution m/z value* | 27.0234; $C_2H_3^+$ |
| *Calibration Species* | $CH_3^+$, $C_4H_5^+$, $C_3H_6NO_2^+$, $C_8H_{10}NO^+$ |
| *Primary Ion Gun* | Liquid Metal Ion Gun |
| *Primary Ion Source* | $Bi_3^+$ |
| *Primary Ion Dose* | $2.45 \times 10^{11}$ ion/cm$^2$ |
| *Pulse Width* | 25.0 ns |
| *Pulse Rate* | 10.0 kHz |
| *Beam Diameter* | 5 μm |
| *Analysis Area* | 500 μm × 500 μm |
| *Polarity* | Positive |
| *Mass Range* | 0-800 |
| *Spectra Acquisition Time* | 40.96 s |